\documentclass{article}

\PassOptionsToPackage{numbers,sort&compress}{natbib}
\usepackage[dblblindworkshop, final]{neurips_2026}
\workshoptitle{Sim2Science: ML with Imperfect Scientific Models}

\usepackage[utf8]{inputenc} 
\usepackage[T1]{fontenc}    
\usepackage{hyperref}       
\usepackage{url}            
\usepackage{booktabs}       
\usepackage{amsfonts}       
\usepackage{nicefrac}       
\usepackage{microtype}      
\usepackage{xcolor}         
\usepackage{graphicx}
\usepackage{amsmath}
\usepackage{amssymb}
\usepackage{capt-of}
\usepackage{float}
\usepackage{orcidlink}
\usepackage{wrapfig}

\ifdefined\anontrue\else
  \expandafter\newif\csname ifanon\endcsname
  \anonfalse
\fi

\providecommand{\mbh}{M_\mathrm{BH}}
\providecommand{\msun}{\mathrm{M}_\odot}
\providecommand{\dex}{\,\mathrm{dex}}
\providecommand{\Mopen}{$\mathcal{M}$-open}

\providecommand{\bd}{\tilde{\mathbf{d}}}
\providecommand{\bm}[1]{\tilde{\mathbf{m}}_{#1}}
\providecommand{\ba}{\mathbf{a}}
\providecommand{\bdelta}{\tilde{\boldsymbol{\delta}}}
\providecommand{\bnoise}{\tilde{\boldsymbol{\eta}}}
\providecommand{\btheta}{\boldsymbol{\theta}}

\providecommand{\arcsec}{\ensuremath{^{\prime\prime}}}

\anonfalse

\ifanon

  \providecommand{\companionp}{\citep{Anon_companion}}
  \providecommand{\companiont}{the companion paper \citep{Anon_companion}}
  
\else

  \providecommand{\companionp}{\citep{YB_2026}}
  \providecommand{\companiont}{\citet{YB_2026}}
  
\fi

\providecommand{\answerTODO}[1][]{\textcolor{red}{\bf [TODO]}}

\title{When precision exceeds fidelity: systematic error quantification for inference with misspecified simulators}

\author{%
    \textbf{M. J. Yantovski-Barth}$^{1,2,3,4}$\thanks{E-mail correspondence: michael.barth@umontreal.ca}\orcidlink{0000-0001-5200-4095} \and 
    \textbf{Hengyue Zhang}$^{5}$\orcidlink{0000-0003-4079-2447} \and 
    \textbf{Nolan Smyth}$^{1,2,3}$\orcidlink{0000-0002-8454-3015} \and
    \textbf{Aristeidis Amvrosiadis}$^{4}$\orcidlink{0000-0002-4465-1564}\and 
    \textbf{Martin Bureau}$^{4}$\orcidlink{0000-0003-4980-1012} \and 
    \textbf{Yashar Hezaveh}$^{1,2,3,6}$\orcidlink{0000-0002-8669-5733} \and 
    \textbf{Laurence Perreault-Levasseur}$^{1,2,3,6}$\orcidlink{0000-0003-3544-3939}
    \\
    $^{1}$Universit{\'e} de Montr{\'e}al \quad $^{2}$Ciela Institute \quad $^{3}$Mila \quad $^{4}$University of Oxford \quad \\ $^{5}$Max Planck Institute for Astronomy \quad $^{6}$CCA, Flatiron Institute
}
\begin{document}
\maketitle
\begin{abstract}
  As scientific datasets grow in complexity and precision, the primary source of uncertainty in parameter inference shifts from measurement noise to systematic errors within the simulators themselves. In this setting, Bayesian evidence concentrates on whichever misspecified model best approximates the data. Evidence-weighted model averaging can then underestimate systematic uncertainty when one model component compensates for the misspecification of another component. We posit that in this case, the key question is how each model responds to discrepancies between simulator and data. We show that the local response of an inferred parameter to perturbations in the data --- the model's influence map --- can be estimated efficiently from forward evaluations of the simulator at existing posterior samples. Propagating discrepancy residuals through these influence maps enables individual sources of systematic bias to be isolated. Applied to independent observations used to infer the mass of a supermassive black hole, we find that the error is dominated by model systematics. We identify a simplification shared by all models, propagate the discrepancies through each model's influence map, and apply a first-order correction of the resulting biases. This substantially improves the cross-observation consistency of the inferred black-hole mass within each model while preserving the between-model spread that represents remaining systematics.
\end{abstract}

\section{Introduction}
\label{sec:intro}

Many scientific applications rely on parameter inference in which a forward model $f(\btheta)$ is combined with a noise model into a likelihood, whether that likelihood is evaluated explicitly, as we consider here, or learned from simulations as in simulation-based inference \citep[e.g.,][]{Cranmer_2020}. Every simulator is a simplification, and growing datasets raise the stakes of simulator misspecification: posterior widths shrink with the number of independent measurements $N$ while the simplifications built into the models remain. In our application ($N\sim10^{5}$--$10^{6}$ independent constraints) the statistical uncertainty on our parameter of interest is 2--20 times smaller than the dispersion across our candidate models.

Evidence-weighted Bayesian model averaging (BMA) \citep{Hoeting_1999} is not designed to treat a misspecified set of simulators as an empirical systematic-error ensemble. In the \Mopen{} setting, posterior model weights concentrate on the candidate closest to the data-generating distribution in Kullback--Leibler divergence \citep{Bernardo_1994,Yao_2018}. The between-model component of the BMA variance can often become negligible even though the selected candidate remains misspecified and its physical parameter estimate may be biased due to compensation between various components of the simulator.

This shifts the question from identifying the best-fitting candidate to quantifying how each candidate responds to discrepancies between simulator and data. We characterize this response using influence functions \citep{Hampel_1974,Cook_1986,Koh_2017}. For a Gaussian likelihood, the derivative of a posterior mean with respect to the whitened data---which we call the model's \emph{influence map}---is a posterior covariance between the parameter of interest and the whitened simulator output \citep{Giordano_2018}. It can therefore be estimated from forward evaluations at existing posterior samples, without simulator gradients or inversion of a Fisher matrix. We contribute (i) a sample-based influence-map estimator and a variance-consistency diagnostic comparing locally propagated noise variance with posterior variance; (ii) a first-order separation of noise-induced and persistent between-model scatter, together with the null-calibrated propagation of an externally estimated discrepancy into model-specific parameter shifts; and (iii) an application to a supermassive black hole (SMBH) mass inferred from multi-resolution ALMA data.

\section{Influence maps for an ensemble of misspecified simulators}
\label{sec:influence}

\paragraph{Setting.}
Let $\mathbf{d}\in\mathbb{R}^{N}$ be the data and $f^{(k)}(\btheta_k)$ the simulator of candidate model $k$ with parameters $\btheta_k$. With independent Gaussian noise of known standard deviations $\sigma_i$, we write $\bd$ for the whitened data ($\tilde d_i = d_i/\sigma_i$) and $\bm{k}(\btheta_k) = f^{(k)}(\btheta_k)/\boldsymbol{\sigma}$ for the whitened simulator output, so that $\ln p(\mathbf{d}\mid\btheta_k) = -\tfrac{1}{2}\lVert\bd - \bm{k}(\btheta_k)\rVert^{2} + \mathrm{const.}$ Nested sampling yields the posterior $p(\btheta_k\mid\mathbf{d})$ and the evidence $Z_k$. For a scalar quantity of interest $\theta$ (here $\theta=\log_{10}\mbh$), $\mu_k$ and $\sigma_k^2$ denote its posterior mean and variance under model $k$. With BMA weights $w_k = Z_k/\sum_j Z_j$ and $\bar\mu=\sum_k w_k\mu_k$, the law of total variance gives
\begin{equation}\label{eq:total_variance}
    \mathrm{Var}\!\left[\theta \mid \mathbf{d}\right]
    = \underbrace{\sum_{k} w_k\,\sigma_k^2}_{\text{statistical}}
    + \underbrace{\sum_{k} w_k \left(\mu_k - \bar{\mu}\right)^2}_{\text{systematic}}\,.
\end{equation}
In the \Mopen{} regime the differences in $\ln Z_k$ grow linearly with $N$: the weights asymptotically collapse onto a single $k$, and the systematic term becomes zero. Predictive stacking \citep{Yao_2018} instead chooses weights to optimize held-out predictive performance. Although stacking is well suited to prediction in the \Mopen{} setting, predictive fit also rewards biases introduced via compensation by multiple model components. A misspecification in one component of the simulator can be compensated by whichever other component is flexible enough to absorb it; if that component is degenerate with the parameter of interest, the compensation biases the estimate while improving the fit. We therefore do not weight by fit at all, but use equal weights, $w_k=1/K_g$, within families $g$ of $K_g$ models whose members use the same external information. The resulting between-model dispersion measures how much the estimate depends on the choice of model within the ensemble; the equal weights express no preference among models and are not posterior model probabilities. The two terms of Eq.~\eqref{eq:total_variance} then become the statistical variance $\sigma_{\rm stat}^2 = \frac{1}{K_g}\sum_{k\in g}\sigma_k^2$ and the observed spread $S_g^2 = \frac{1}{K_g}\sum_{k\in g}(\mu_k-\bar{\mu}_g)^2$ about the family mean $\bar{\mu}_g$. The influence map methodology described below helps us interpret the observed spread.

\paragraph{Definition and estimation from posterior samples.}
The influence map of model $k$ is the derivative of $\mu_k$ with respect to the whitened data. For any likelihood, $\partial_{\mathbf{d}}\mu_k = \mathrm{Cov}_\mathrm{post}[\theta,\partial_{\mathbf{d}}\ln p(\mathbf{d}\mid\btheta_k)]$ \citep{Giordano_2018,Giordano_2023}; for the Gaussian likelihood $\partial_{\bd}\ln p(\mathbf{d}\mid\btheta_k) = \bm{k}(\btheta_k)-\bd$ (see Appendix \ref{app:posterior_covariance_identity} for a full derivation). The whitened data ($\bd$) is a constant and thus uncorrelated with the parameters $\btheta_k$, leaving
\begin{equation}\label{eq:influence}
    \ba_k
    \equiv \frac{\partial \mu_k}{\partial \bd}
    = \mathrm{Cov}_{\mathrm{post}}\!\left[\,\theta,\;
      \bm{k}(\btheta_k)\,\right],
\end{equation}
the posterior covariance between $\theta$ and the whitened simulator output over parameter vectors $\btheta_k$ drawn from the posterior of model $k$. This identity is exact for both linear and nonlinear simulators with fixed Gaussian noise and a data-independent prior. To first order, the parameter of interest then responds to a whitened data perturbation $\bnoise$---which could be a noise realization or a deterministic misspecification residual---as $\Delta\mu_k = \ba_k^{\top}\bnoise$. In the linear-Gaussian, flat-prior limit Eq.~\eqref{eq:influence} reduces to the Fisher form $\ba_k = \tilde{J}_k(\tilde{J}_k^{\top}\tilde{J}_k)^{-1}\mathbf{e}$ \citep{Flanagan_1998,Cutler_2007,Amara_2008}, with $\tilde{J}_k$ the whitened Jacobian and $\mathbf{e}$ selecting $\theta$. However, this latter form requires an invertible Fisher matrix. For the flexible models we use in Section~\ref{sec:application}, the Fisher matrix is near-singular along degeneracies which are broken only by the prior and by posterior curvature. Thus, we adopt Eq.~\eqref{eq:influence} throughout and estimate it via Monte Carlo over $n_s = 512$ equal-weight posterior draws $\btheta_{k,s}$, one simulator evaluation each:
\begin{equation}\label{eq:influence_mc}
    \hat{\ba}_k
    = \frac{1}{n_s}\sum_{s=1}^{n_s}
      \big(\theta_{k,s}-\bar{\theta}_k\big)
      \big(\tilde{\mathbf{m}}_{k,s}-\bar{\tilde{\mathbf{m}}}_k\big)\,,
\end{equation}
with $\theta_{k,s}$ the value of $\theta$ at draw $s$, $\tilde{\mathbf{m}}_{k,s}=\bm{k}(\btheta_{k,s})$ and bars denoting means over draws.

\paragraph{Monte Carlo error and the consistency ratio.}
Linear functionals of $\hat{\ba}_k$ are unbiased by its Monte Carlo scatter and carry a standard error from the per-draw terms; quadratic functionals are biased upward by $\sum_i \mathrm{Var}\,\hat{a}_{k,i}$ and are debiased with per-component variances from the same draws (Appendix~\ref{app:mc}). We also compute from the debiased norm the \emph{variance-consistency ratio} $\rho_k\equiv\lVert\ba_k\rVert^2/\sigma_k^2$, which compares the locally propagated noise variance of the posterior mean with the posterior variance. Equality is expected only under likelihood-dominated linear-Gaussian conditions, so $\rho_k$ serves as a consistency diagnostic for our influence map-derived functionals.

\paragraph{Decomposing the ensemble spread.}
To separate persistent between-model differences from variation induced by the particular noise realization, we first introduce $\mu_k^{\infty}$, the theoretical parameter mean which we would measure if we observed noiseless data (a practical impossibility). To first order, $\mu_k=\mu_k^{\infty}+\ba_k^\top\bnoise$, where $\bnoise$ is a unit-covariance noise realization. Substituting into $S_g^2$ and averaging over $\bnoise$,
\begin{equation}\label{eq:spread_decomp}
    \mathbb{E}\!\left[S_g^2\right]
    = \underbrace{\frac{1}{K_g}\sum_{k\in g}
      \big(\mu_k^{\infty} - \bar{\mu}_g^{\infty}\big)^2}_{\text{persistent between-model}}
    + \underbrace{\frac{1}{K_g}\sum_{k\in g}
      \big\lVert \ba_k - \bar{\ba}_g \big\rVert^2}_{\text{noise-induced}}\,,
\end{equation}
with $\bar{\ba}_g$ the family-mean influence map. The noise-induced term follows from the influence maps alone (Appendix~\ref{app:mc}), which allows us to calculate $\hat{S}^2_{g,\mathrm{noise}}$ and the persistent between-model spread $\hat{S}^2_{g,\mathrm{sys}} = S_g^2 - \hat{S}^2_{g,\mathrm{noise}}$. We note here that a shared misspecification residual $\bdelta$ biases the family mean through $\bar{\ba}_g^{\top}\bdelta$, to which the spread is blind, and leaks into the apparent model spread through $(\ba_k-\bar{\ba}_g)^{\top}\bdelta$; quantifying either requires estimating $\bdelta$ explicitly, as follows.

\paragraph{Propagating an externally estimated discrepancy residual.}
In order to quantify the effect of a single model misspecification, we fix the parameters of interest to their best-fit values and re-fit a more flexible parametrization of the misspecified model component \citep[a modular or \emph{cut} construction:][]{Plummer_2015,Jacob_2017}. From there, we calculate how much the resulting shift in predicted model could have biased the parameters of interest. For a deterministic whitened residual $\bdelta$, each model's bias is $\Delta\hat{\mu}_k = \ba_k^{\top}\bdelta$ to first order. In our application, the refitted pixelated (nonparametric) component $\mathbf f$ enters model $k$ linearly at fixed remaining parameters,
$\mathbf m_k=A_k\mathbf f$. Starting from the fitted parametric profile $\mathbf f_{\mathrm{param},k}$, we solve
$\mathbf f_{\mathrm{opt},k}
=\operatorname{arg\,min}_{\mathbf f\ge0}
\lVert\bd-\tilde A_k\mathbf f\rVert^2$
while holding the remaining parameters at $\hat{\btheta}_k$. Non-negativity stabilizes the inversion (see Appendix \ref{app:refit}). We define the model-specific data-space discrepancy
$\bdelta_k=\tilde A_k(\mathbf f_{\mathrm{opt},k}-\mathbf f_{\mathrm{exp},k})$
and estimate the corresponding shift in the parameter of interest as
$\Delta\hat\mu_k=\ba_k^\top\bdelta_k$.
Because the refit acts nonlinearly on noise, we calibrate this statistic using 200 parametric Monte Carlo mocks \citep{Hope_1968,Gelman_1996}. We adopt a mock noise scaling
$s_k=\sqrt{\chi^2(\hat{\btheta}_k)/N}$
to match the initial reduced $\chi^2$ of the mocks to that of the observed fit; we then optimize $\mathbf f$ and calculate the data-space discrepancy in the same way as above. We then estimate the bias in the parameter of interest as 
$\Delta\hat\mu_k-\langle\Delta\mu\rangle_{\mathrm{null}}$, where $\langle\Delta\mu\rangle_{\mathrm{null}}$ is the average bias among the mocks.

\section{Application: a black hole mass from multi-resolution ALMA data}
\label{sec:application}

The mass of a supermassive black hole can be inferred from molecular-gas kinematics within its gravitational sphere of influence \citep{Davis_2013b}. Here the gas is traced by CO(2--1) emission observed with the Atacama Large Millimeter Array (ALMA) in two independent observations of the elliptical galaxy NGC~4697 \citep{Davis_2017}, at lower ($\simeq0.5''$) and higher ($\simeq0.065''$) angular resolution. We analyse the observations separately and jointly (referred to as the ``combined'' dataset). Interferometry data from telescopes such as ALMA are taken in Fourier space. In the Fourier domain, the thermal-noise contribution is approximately Gaussian; our non-overlapping data gridding procedure ensures no overlap-induced covariance between cells \companionp{}. The three datasets contain $N=1.0\times10^5 - 1.6\times10^6$ real-valued likelihood terms. Our simulator, \texttt{SuperMAGE} \companionp{} renders a thin circular gas disc whose velocity is set by a stellar mass profile and a central point mass $\mbh$, then transforms the resulting spectral cube into model Fourier-plane data (see Appendix~\ref{app:settings} for all simulator settings). All models use an exponential gas-emissivity profile \citep{Davis_2013b} as a common baseline approximation. We compare two imaging-based stellar profiles, which differ in whether the innermost AGN-contaminated MGE Gaussian is retained (MGE+AGN) or removed (MGE), with three imaging-free profiles (S\'ersic, Nuker, and Core-S\'ersic) inferred from the kinematics. This comparison tests whether imaging-free models recover black hole masses consistent with conventional imaging-anchored models.

\begin{wraptable}{r}{0.52\linewidth}
    \centering
    \small
    \setlength{\tabcolsep}{3pt}
    \caption{Uncertainty budget of $\log_{10}\mbh$ (dex) per dataset and model family: mean within-model posterior width $\sigma_\mathrm{stat}$, noise-induced between-model spread $S_\mathrm{noise}$, and persistent non-noise spread $S_\mathrm{sys}$, Eq.~\eqref{eq:spread_decomp}; $S_\mathrm{obs}^2=S_\mathrm{sys}^2+S_\mathrm{noise}^2$.}
    \label{tab:budget}
    \begin{tabular}{ll ccc}
      \toprule
      Dataset   & Family        & $\sigma_\mathrm{stat}$ & $S_\mathrm{noise}$ & $S_\mathrm{sys}$ \\
      \midrule
      lower-res & imaging-based & 0.007 & 0.002 & 0.085 \\
                & imaging-free  & 0.052 & 0.026 & 0.101 \\
      high-res  & imaging-based & 0.005 & 0.001 & 0.082 \\
                & imaging-free  & 0.018 & 0.009 & 0.081 \\
      combined  & imaging-based & 0.004 & 0.001 & 0.082 \\
                & imaging-free  & 0.022 & 0.016 & 0.084 \\
      \bottomrule
    \end{tabular}
\end{wraptable}

\paragraph{Model dependence dominates the statistical error budget.}
Across every dataset and model family, the observed between-model spread of $0.08$--$0.10\dex$ exceeds the mean within-model posterior width of $0.004$--$0.052\dex$, by factors of 2--20 (see Table \ref{tab:budget}). Differential responses to the realized noise contribute $\leq 0.026\dex$ to the between-model spread, so most of the model dependence persists after accounting for noise. Bayesian evidence, meanwhile, strongly favours different models at different angular resolutions (see Appendix \ref{app:evidences}).

\begin{table}[b]
  \caption{Null-calibrated emissivity-induced bias of $\log_{10}\mbh$, $\Delta\hat{\mu}_k - \langle\Delta\mu\rangle_\mathrm{null}$ (dex), $\pm$ the standard deviation of the 200-mock null distribution; this is the quantity subtracted in Figure~\ref{fig:corrected_bh_mass_comparison}. Uncalibrated values, null means and Monte Carlo standard errors ($\le0.03\dex$) are listed in Appendix~\ref{app:refit}. Bold: more than two null standard deviations from zero.}
  \label{tab:flux_bias}
  \centering
  \small
  \begin{tabular}{l ccc}
    \toprule
    Model         & lower-res & high-res & combined \\
    \midrule
    MGE           & $+0.001\pm0.005$ & $+0.001\pm0.004$ & $\phantom{+}0.000\pm0.003$ \\
    MGE+AGN       & $\phantom{+}0.000\pm0.007$ & $-0.001\pm0.005$ & $-0.001\pm0.004$ \\
    S\'ersic      & $\mathbf{+0.221\pm0.047}$ & $+0.009\pm0.012$ & $+0.011\pm0.010$ \\
    Nuker         & $\mathbf{+0.094\pm0.035}$ & $+0.004\pm0.018$ & $+0.012\pm0.014$ \\
    Core-S\'ersic & $\mathbf{+0.086\pm0.035}$ & $+0.013\pm0.019$ & $-0.051\pm0.029$ \\
    \bottomrule
  \end{tabular}
\end{table}

\paragraph{Flexible stellar profiles compensate for emissivity misspecification at low resolution.}
We estimate the effect of the shared exponential-emissivity approximation by replacing it with a non-negative pixelated refit under the cut construction of Section~\ref{sec:influence} (visualized in Appendix~\ref{app:figures}). Propagating the resulting data-space discrepancies through the influence maps gives estimated black-hole-mass biases that depend strongly on both the stellar parametrization and the angular resolution (Table~\ref{tab:flux_bias}). The shifts are $\leq0.001\dex$ and consistent with the null for the imaging-based models in every dataset. They are also consistent with the null for the imaging-free models fitted to the high-resolution and combined data. In the lower-resolution data, however, all three imaging-free models bias $\log_{10}\mbh$ high by $0.086$--$0.221\dex$, corresponding to 2.5--4.7 standard deviations of their null distributions.

We interpret this bias pattern as compensation for the misspecification of the exponential emissivity model through the stellar mass component. Misspecification of the gas-emissivity distribution can be partly accommodated by changing the inferred stellar profile. At lower resolution, the inner part of an imaging-free stellar profile is inferred from the same kinematics as $\mbh$, so the stellar mass in the unresolved central region remains degenerate with the black hole. Thus, a slight change in the stellar mass profile can significantly affect its extrapolation into the unresolved central region and thereby change the partition of the enclosed mass between stars and the black hole. This results in a biased black-hole mass while improving the fit to the lower-resolution data. The externally constrained imaging-based profiles restrict this compensation route, while the high-resolution observation better separates the dynamical signature of the central black-hole from the surrounding stellar mass. 

\begin{figure}[t]
\centering
\includegraphics[width=0.45\linewidth]{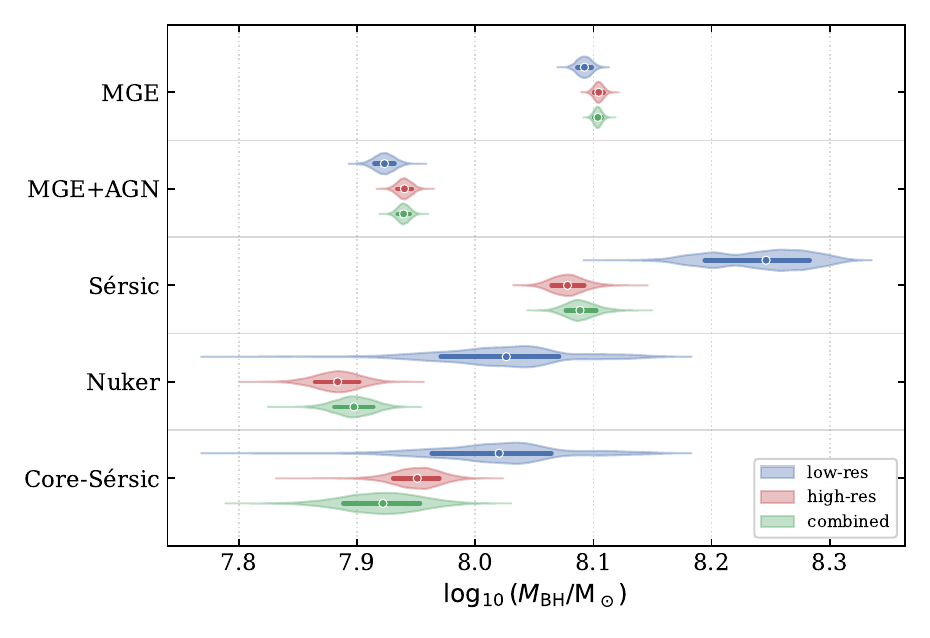}\hfill
\includegraphics[width=0.45\linewidth]{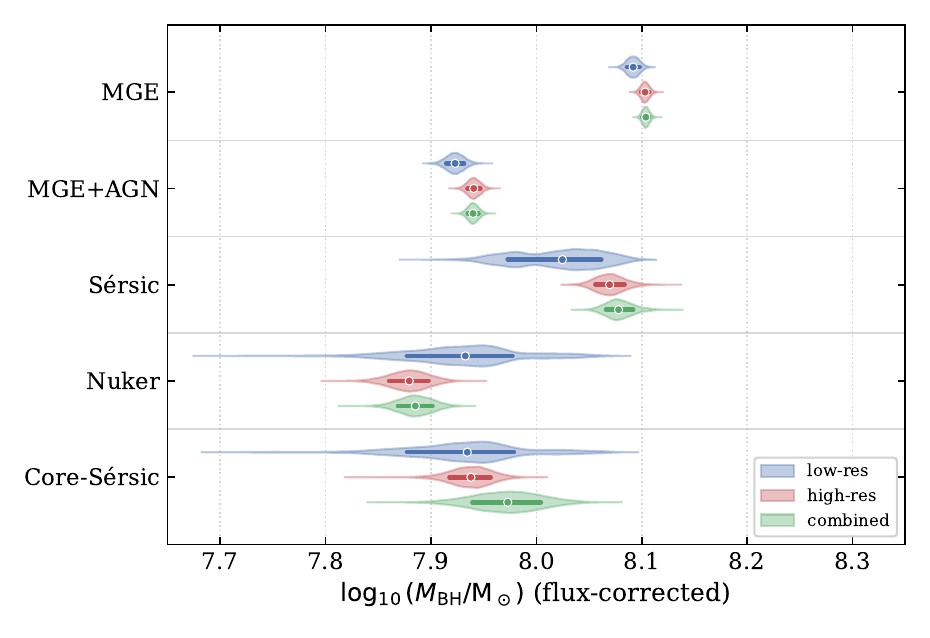}
\caption{Posterior distributions of $\log_{10}(\mbh/\msun)$ before (left) and after translation by the null-calibrated first-order emissivity shifts (right). Error bars show the 15.86th and 84.14th percentiles and exclude uncertainty in the estimated shifts.}
\label{fig:corrected_bh_mass_comparison}
\vspace{-1.2em}
\end{figure}

\paragraph{After removing the coherent bias.}
Subtracting the null-calibrated biases from the measured black hole mass brings each imaging-free model into statistical agreement with itself across datasets (Figure~\ref{fig:corrected_bh_mass_comparison}) by correcting the discrepant lower-resolution masses. The variance-consistency ratios are $0.93$--$1.44$ except for the lower-resolution S\'ersic fit, for which $\rho_k=3.3$ (for a full list, see Table~\ref{tab:flux_bias_full}), so the correction to the S\'ersic model's black hole mass may be unreliable. Another caveat is that we do not propagate the uncertainty in the shifts into the resulting credible intervals. For the high-resolution and combined datasets, the $\approx0.08$ dex between-model spread that remains after correction exceeds the statistical uncertainty by factors of 3–20 (for all corrected spreads, see Table~\ref{tab:systematic_uncertainties}). For the lower-resolution imaging-free family the correction reduces the spread from 0.10 to $0.03$ dex, below its statistical width of 0.05 dex, so model systematics are no longer dominant there. Using Eq. \eqref{eq:total_variance}, for the combined dataset, our corrected black hole mass measurement yields $\log_{10}(\mbh/\msun)=7.98\pm0.02\,(\mathrm{stat})\pm0.08\,(\mathrm{sys})$ for the imaging-free models and $8.02\pm0.004\pm0.08$ for the imaging-based models. Thus, the two different types of stellar mass models lead to black hole masses that are consistent to within the measured systematic uncertainty. 

\section{Discussion and conclusions}
\label{sec:discussion}

Our application illustrates a consequential failure mode of inference with misspecified simulators: one model component can compensate for an error in another, improving the fit while biasing the physical parameter of interest. In the lower-resolution observation, the imaging-free models compensate for the shared exponential-emissivity approximation through changes in the inferred inner stellar mass profile. Because the kinematics constrain the combined enclosed mass of the stars and black hole, this compensation biases $\log_{10}\mbh$. The effect is suppressed when the stellar profile is constrained externally by imaging or when the central kinematics are observed at higher angular resolution. Propagating the emissivity discrepancies through model-specific influence maps substantially improves the cross-observation consistency of every imaging-free model.

The influence maps and spread decomposition require $n_s=512$ simulator evaluations per model and dataset (compared with approximately $10^6$ evaluations for the original inference) and require neither simulator gradients nor an invertible Fisher matrix. They can therefore be evaluated for black-box simulators, prior-regularized posteriors, and amortized posteriors. Propagating a particular discrepancy additionally requires a flexible representation of the suspected component, a modular refit that prevents feedback into the scientific parameters, and null calibration of the resulting statistic.

The principal limitations are the first-order and modular nature of the calculation. The variance-consistency ratio $\rho_k$ is sensitive to posterior geometry in addition to nonlinear conditions. Thus, neither $\rho_k\simeq1$ validates a finite first-order shift nor $\rho_k>1$ uniquely diagnoses nonlinearity. Large discrepancy shifts should ultimately be checked by re-inference in which the flexible component is marginalized \citep{Suyu_2006} or represented in a discrepancy-aware likelihood \citep{Kennedy_2001,Brynjarsdttir_2014}. The ensemble spread is also conditional on the selected parametrizations; it is therefore not a complete systematic uncertainty. Here we propagate the gas-emissivity approximation but not other shared assumptions such as circular motion in a thin disc. Future work will also investigate improved consistency tests using simulated observations and apply the influence map methodology to additional science cases.

\section{Acknowledgements}
This work is partially supported by Schmidt Sciences, a philanthropic initiative founded by Eric and Wendy Schmidt as part of the Virtual Institute for Astrophysics (VIA). This work is in part supported by computational resources provided by Calcul Quebec and the Digital Research Alliance of Canada. 
M.J.Y.B. acknowledges support from Mitacs through the Globalink program and from the Balzan Centre for Cosmological Studies at New College, University of Oxford via the Balzan Fellowship. M.J.Y.B also acknowledges support from UdeM through the Bourse d’excellence du centenaire, Bourse du passage accéléré, and the Bourse Gilles Beaudet. M.J.Y.B acknowledges technical support from Claude Pascal Duquette and Luc Turbide.
H.Z. acknowledges support from a Science and Technology Facilities Council (STFC) DPhil studentship under grant ST/X508664/1 and the Balliol College J. T. Hamilton Scholarship in physics. M.B. was supported by STFC consolidated grant ‘Astrophysics at Oxford’ ST/K00106X/1 and ST/W000903/1.
Y.H. and L.P. acknowledge support from the Canada Research Chairs Program, the Natural Sciences and Engineering Research Council of Canada through grants RGPIN-2020-05073 and 05102.

\bibliographystyle{plainnat}
\bibliography{refs}

\newpage

\appendix
\section{Derivation of the posterior-covariance identity}
\label{app:posterior_covariance_identity}

Let $\boldsymbol\theta_k$ denote the parameters of model $k$, and let
$\theta$ be the scalar parameter of interest. The posterior distribution
and its mean are
\begin{equation}
    p_k(\boldsymbol\theta_k\mid\tilde{\mathbf d})
    = \frac{L_k(\tilde{\mathbf d}\mid\boldsymbol\theta_k)
    \,\pi_k(\boldsymbol\theta_k)}{Z_k(\tilde{\mathbf d})},
    \qquad
    \mu_k = \int \theta\,
    p_k(\boldsymbol\theta_k\mid\tilde{\mathbf d})
    \,\mathrm d\boldsymbol\theta_k,
\end{equation}
where
\begin{equation}
    Z_k(\tilde{\mathbf d})
    = \int L_k(\tilde{\mathbf d}\mid\boldsymbol\theta_k)
    \,\pi_k(\boldsymbol\theta_k)\,\mathrm d\boldsymbol\theta_k.
\end{equation}
We assume that the prior, parameter support, and quantity of interest
are independent of the data, that the whitening transformation is held
fixed, and that differentiation may be interchanged with integration.
All expectations and covariances below are taken under the posterior
of model $k$ at the observed data.

Define the likelihood score with respect to the data as
\begin{equation}
    \mathbf s_k
    \equiv \nabla_{\tilde{\mathbf d}}\log L_k.
\end{equation}
Since $\nabla_{\tilde{\mathbf d}}L_k=L_k\mathbf s_k$,
differentiating the normalization gives
\begin{equation}
    \nabla_{\tilde{\mathbf d}}Z_k
    = \int L_k\,\pi_k\,\mathbf s_k
    \,\mathrm d\boldsymbol\theta_k
    = Z_k\,\mathbb E_{\mathrm{post}}[\mathbf s_k].
\end{equation}
Consequently, the derivative of the normalized posterior is
\begin{equation}
\begin{aligned}
    \nabla_{\tilde{\mathbf d}}p_k
    &= \frac{\pi_k\,\nabla_{\tilde{\mathbf d}}L_k}{Z_k}
    - \frac{L_k\pi_k}{Z_k^2}
    \,\nabla_{\tilde{\mathbf d}}Z_k \\
    &= p_k\left(
    \mathbf s_k-\mathbb E_{\mathrm{post}}[\mathbf s_k]
    \right).
\end{aligned}
\end{equation}
Differentiating the posterior mean therefore yields
\begin{equation}
\begin{aligned}
    \mathbf a_k
    \equiv \nabla_{\tilde{\mathbf d}}\mu_k
    &= \int \theta\,\nabla_{\tilde{\mathbf d}}p_k
    \,\mathrm d\boldsymbol\theta_k \\
    &= \mathbb E_{\mathrm{post}}[\theta\,\mathbf s_k]
    - \mathbb E_{\mathrm{post}}[\theta]
    \,\mathbb E_{\mathrm{post}}[\mathbf s_k] \\
    &= \operatorname{Cov}_{\mathrm{post}}
    [\theta,\mathbf s_k].
\end{aligned}
\end{equation}

For the whitened Gaussian likelihood,
\begin{equation}
    \log L_k
    = -\frac12\left\|
    \tilde{\mathbf d}
    -\tilde{\mathbf m}_k(\boldsymbol\theta_k)
    \right\|^2+\mathrm{const},
\end{equation}
the score is
\begin{equation}
    \mathbf s_k
    = \tilde{\mathbf m}_k(\boldsymbol\theta_k)
    -\tilde{\mathbf d}.
\end{equation}
Because the observed data $\tilde{\mathbf d}$ are constant within the
posterior expectation, their covariance with $\theta$ vanishes, giving
\begin{equation}
    \boxed{
    \mathbf a_k
    = \nabla_{\tilde{\mathbf d}}\mu_k
    = \operatorname{Cov}_{\mathrm{post}}
    \!\left[\theta,\tilde{\mathbf m}_k\right].
    }
    \label{eq:posterior_covariance_identity_derivation}
\end{equation}

\section{Monte Carlo error of influence-map functionals}
\label{app:mc}

The influence-map estimate $\hat{\ba}_k$ of Eq.~\eqref{eq:influence_mc} is an average over $n_s=512$ posterior draws and therefore carries Monte Carlo noise, which affects linear and quadratic functionals differently.

\paragraph{Linear functionals.}
Any propagated perturbation $\Delta\mu_k = \ba_k^{\top}\bdelta$ is linear in $\hat{\ba}_k$ and hence unbiased by random scatter in $\hat{\ba}_k$; its Monte Carlo scatter is measured directly by rewriting the estimate as the mean over draws of one scalar per draw,
\begin{equation}\label{eq:mc_se}
    \widehat{\Delta\mu}_k = \frac{1}{n_s}\sum_s c_{k,s}\,,\qquad
    c_{k,s} = \big(\theta_{k,s}-\bar{\theta}_k\big)
          \big(\tilde{\mathbf{m}}_{k,s}^{\top}\bdelta
          - \overline{\tilde{\mathbf{m}}_k^{\top}\bdelta}\big)\,,
\end{equation}
and quoting $\mathrm{SE} = \mathrm{std}(c_{k,s})/\sqrt{n_s}$. Every propagated bias carries this standard error (Table~\ref{tab:flux_bias_full}).

\paragraph{Quadratic functionals.}
Writing $\hat{\ba}_k = \ba_k + \boldsymbol{\varepsilon}$ with $\mathbb{E}[\boldsymbol{\varepsilon}]=0$, and with the index $i=1,\dots,N$ running over the components of the whitened data vector,
\begin{equation}
    \mathbb{E}\,\lVert\hat{\ba}_k\rVert^2
    = \lVert\ba_k\rVert^2
    + \sum_i \mathrm{Var}\,\hat{a}_{k,i}\,,
\end{equation}
because the noise cannot cancel in a square. We estimate the per-component variance from the same draws via the variance-of-a-sample-mean formula,
\begin{equation}\label{eq:variance_ak}
    V_{k,i} \equiv \widehat{\mathrm{Var}}\,\hat{a}_{k,i} =
    \frac{\,\overline{c_{k,s,i}^2} - \hat{a}_{k,i}^2\,}{n_s}\,,
\end{equation}
with $c_{k,s,i}$ the per-draw centred products of Eq.~\eqref{eq:influence_mc}, and obtain the debiased norm
\begin{equation}\label{eq:norm_debias}
    \lVert\ba_k\rVert^2
    \equiv \lVert\hat{\ba}_k\rVert^2 - \sum_i V_{k,i}\,.
\end{equation}
Debiasing centres the estimator but does not remove its scatter; the residual uncertainty of Eq.~\eqref{eq:norm_debias} is quantified by a jackknife over sixteen 32-draw shards of the $n_s=512$ samples. All products are accumulated in double precision.

\paragraph{Consistency ratio.}
Since the linear response $\Delta\mu_k = \ba_k^{\top}\bnoise$ (Section~\ref{sec:influence}) is only accurate to first order, a large propagated shift is likely to be affected by higher-order effects. The consistency ratio,
\begin{equation}\label{eq:consistency}
    \rho_k \equiv
    \frac{\lVert\ba_k\rVert^2}{\sigma_{k}^2}
    = \frac{\mathrm{Var}_\mathrm{noise}(\mu_k)}{\mathrm{Var}_\mathrm{post}(\theta)}\,,
\end{equation}
is the ratio of the noise-induced variance of the reported mean to the posterior variance. In the linear, data-dominated limit $\rho_k\lesssim1$; we use $\rho_k$, per model and dataset (Table~\ref{tab:flux_bias_full}), to delimit where the linear propagation is more trustworthy.

\paragraph{Noise-induced spread.}
The same inflation affects the noise-induced term of Eq.~\eqref{eq:spread_decomp}. Because the Monte Carlo draws are independent across models, the inflation of $\frac{1}{K_g}\sum_{k\in g}\lVert\hat{\ba}_k-\hat{\bar{\ba}}_g\rVert^2$ is $(1-1/K_g)\bar{V}_g$, where $\bar{V}_g=\frac{1}{K_g}\sum_{k\in g}V_k$ is the family mean of the per-model totals $V_k=\sum_i V_{k,i}$. We subtract it (clipping at zero),
\begin{equation}\label{eq:noise_induced_debias}
    \hat{S}^2_{g,\mathrm{noise}}
    = \frac{1}{K_g}\sum_{k\in g}
      \big\lVert \hat{\ba}_k - \hat{\bar{\ba}}_g \big\rVert^2
    - \Big(1-\tfrac{1}{K_g}\Big)\bar{V}_g\,,
\end{equation}
and obtain the genuine parametrization systematic as
\begin{equation}
    \hat{S}^2_{g,\mathrm{sys}} = \frac{1}{K_g}\sum_{k\in g}
      \left(\mu_k - \bar{\mu}_g\right)^2 - \hat{S}^2_{g,\mathrm{noise}}\,.
\end{equation}
These are the quantities $S_\mathrm{noise}$ and $S_\mathrm{sys}$ of Table~\ref{tab:systematic_uncertainties}. In practice the constraints on $\mbh$ come from the same high-velocity channels regardless of the model, so the differential noise-induced shift in $\mbh$ is significantly smaller than the noise-driven shift shared by all models within a family.

\paragraph{Cross-check against the Fisher form.}
For the two imaging-based models, whose dynamical parameters are likelihood-driven and whose Fisher matrices are well conditioned, the whitened Jacobians can be computed exactly by forward-mode automatic differentiation through the simulator, and the Fisher form $\ba_k = \tilde{J}_k(\tilde{J}_k^{\top}\tilde{J}_k)^{-1}\mathbf{e}$ provides an independent estimate of the influence map. It matches Eq.~\eqref{eq:influence_mc} at cosine similarity $0.87$--$0.88$ and under-predicts the overall amplitude $\lVert\ba_k\rVert$ by $\sim35$ per cent, reflecting non-Gaussian posterior structure that the linearized form omits; the differences between the two estimates produce a negligible ($\lesssim0.003\dex$) shift in every functional quoted in the main text. For the imaging-free profiles the Fisher form is numerically intractable, because the degeneracies between $\mbh$ and the inner stellar profile are regularized by the priors and by the curvature of the degeneracy rather than by the unregularized likelihood \citep{Vallisneri_2008}; Eq.~\eqref{eq:influence} is therefore adopted as the definition throughout.

\section{Non-negative emissivity refit, null calibration and consistency ratios}
\label{app:refit}

\paragraph{Solver.}
Holding all kinematic and stellar parameters of model $k$ at their best-fitting values, the model visibilities are linear in the sky-plane emissivity, $\mathbf{m}=A\,\mathbf{f}$, and we solve
\begin{equation}\label{eq:flux_refit}
    \mathbf{f}_\mathrm{opt}
    = \operatorname*{arg\,min}_{\mathbf{f}\,\ge\,0}\;
    \big\lVert \bd - \tilde{A}\,\mathbf{f} \big\rVert^{2}
\end{equation}
on the pixel grid of the simulator, initialized at the best-fitting exponential emissivity. At the constrained optimum the Karush--Kuhn--Tucker conditions hold pixel by pixel: where $\mathbf{f}_\mathrm{opt}>0$ the $\chi^2$ gradient vanishes, and where $\mathbf{f}_\mathrm{opt}=0$ it is non-negative, i.e.\ any further reduction of $\chi^2$ at that pixel would require negative flux. Unconstrained, $\tilde{A}$ has a large near-null space at sub-resolution scales and the minimizer would fit noise with oscillating positive and negative flux; non-negativity is what makes the refit stable. We solve Eq.~\eqref{eq:flux_refit} by conjugate-gradient iteration on the normal equations. A step that would drive a pixel negative is clipped at the boundary; that pixel joins the active set and the conjugate directions restart. A projected-gradient fallback completes the solve when repeated restarts stall the conjugate iteration. Iterations terminate at the discrepancy target $\chi^2\le N$, at a KKT point, or---in practice, since the data's excess noise power is not flux-driven and the discrepancy target need not be attainable---at a $\chi^2$ plateau. The identical stopping rule is applied to the observed data and to every null-calibration mock.

Because the gas is a massless tracer of the potential in this system and the model visibilities are linear in the sky-plane emissivity at fixed kinematics, the refit can target the emissivity misspecification without altering the dynamical model. The modular construction \citep{Plummer_2015,Jacob_2017} prevents the refit from feeding back into the kinematic parameters, and its effect on $\mbh$ is instead propagated through the influence map, evaluated per posterior draw (Eq.~\ref{eq:mc_se}) so that every quoted bias carries its Monte Carlo standard error.

\paragraph{Null calibration.}
We generate $n_\mathrm{mock} = 200$ mock datasets from the best-fitting model visibilities,
\begin{equation}\label{eq:mock}
    \bd_\mathrm{mock}
    = \bm{k}(\hat{\btheta}_k)
    + s_k\,\boldsymbol{\epsilon},\qquad
    \boldsymbol{\epsilon}\sim\mathcal{N}(0,\mathbb{I}),\qquad
    s_k = \sqrt{\chi^2(\hat{\btheta}_k)/N},
\end{equation}
each with the same realization $\boldsymbol{\epsilon}$ shared by all models within a mock to preserve the cross-model correlation structure. The per-model noise scaling $s_k$ (a $\sim2$ per cent inflation over the nominal uncertainties) ensures that the optimizer starts as far from the stopping criteria for a mock as it does for the real data. The same operator, initialization, solver and stopping rule are applied to each mock, and each mock records the propagated bias $\hat{\ba}_k^{\top}\bdelta_{\mathrm{mock},j}$ of every model. The observed biases are then contextualized with respect to the null distribution through its mean and standard deviation and through a two-sided empirical $p$-value, whose resolution floor at 200 mocks is $p=0.01$. Because the same $\hat{\ba}_k$ is used for the observed and the mock perturbations, the null-calibrated percentiles are independent of the Monte Carlo noise of the influence maps. Table~\ref{tab:systematic_uncertainties} gives the uncertainty budget and the family means before and after correction, Table~\ref{tab:flux_bias_full} lists the observed biases with their Monte Carlo standard errors, the null means and standard deviations at full precision, and the consistency ratios of Eq.~\eqref{eq:consistency}.

\begin{table}[H]
\centering
\caption{Uncertainty budget of $\log_{10}(\mbh/\msun)$ per dataset and family: statistical width $\sigma_\mathrm{stat}$, observed cross-model spread $S_\mathrm{obs}$, its noise-induced part $S_\mathrm{noise}$ and the genuine parametrization systematic $S_\mathrm{sys}$ (Section~\ref{sec:influence} and Eq.~\ref{eq:spread_decomp}); equal-weight family means before ($\langle m_\mathrm{BH}\rangle$) and after ($\langle m_\mathrm{BH}\rangle_\mathrm{corr}$) removing the null-calibrated emissivity bias of Table~\ref{tab:flux_bias}, and the parametrization systematic after correction, $S_{sys,corr}$, computed from the corrected means with the same $S_{noise}^2$ subtracted. Entries in dex.}
\label{tab:systematic_uncertainties}
\small
\begin{tabular}{ll cccc cc c}
\toprule
Dataset & Family & $\sigma_\mathrm{stat}$ & $S_\mathrm{obs}$ & $S_\mathrm{noise}$ & $S_\mathrm{sys}$ & $\langle m_\mathrm{BH}\rangle$ & $\langle m_\mathrm{BH}\rangle_\mathrm{corr}$ & $S_\mathrm{sys,corr}$ \\
\midrule
lower-res & imaging-based & 0.0073 & 0.0847 & 0.0018 & 0.0847 & 8.008 & 8.007 & 0.0845 \\
lower-res & imaging-free & 0.0524 & 0.1040 & 0.0263 & 0.1006 & 8.093 & 7.960 & 0.0325 \\
high-res & imaging-based & 0.0052 & 0.0822 & 0.0007 & 0.0822 & 8.022 & 8.022 & 0.0812 \\
high-res & imaging-free & 0.0178 & 0.0811 & 0.0093 & 0.0806 & 7.971 & 7.962 & 0.0794 \\
combined & imaging-based & 0.0043 & 0.0822 & 0.0008 & 0.0822 & 8.022 & 8.022 & 0.0819 \\
combined & imaging-free & 0.0223 & 0.0856 & 0.0156 & 0.0842 & 7.969 & 7.979 & 0.0777 \\
\bottomrule
\end{tabular}
\end{table}

\begin{table}[H]
\centering
\caption{Full-precision version of Table~\ref{tab:flux_bias}: per model and dataset, the emissivity-induced bias $\Delta\mu_\mathrm{flux}$ ($d\mu_k\pm\sigma_{\mathrm{MC},k}$, Monte Carlo standard error from Eq.~\ref{eq:mc_se}), the null distribution ($\langle d\mu\rangle_\mathrm{null}\pm\sigma_\mathrm{null}$) obtained by propagating random noise through the same pipeline, and the consistency ratio $\rho_k$ of Eq.~\eqref{eq:consistency}. $d\mu_k$ is the observed black hole mass minus what it would be with the correct emissivity; values of $d\mu_k$ far outside the null indicate significant biases, and $\rho_k$ values significantly above one indicate that the linear approximation may not hold. Biases in dex.}
\label{tab:flux_bias_full}
\small
\begin{tabular}{llccc}
\toprule
Dataset & Model & $\Delta\mu_{\rm flux}$ ($d\mu_k \pm \sigma_{{\rm MC},k}$) & Null ($\langle d\mu\rangle_{\rm null} \pm \sigma_{\rm null}$) & $\rho_k$ \\
\midrule
lower-res & MGE           & $+0.0037\pm0.0012$ & $+0.0032\pm0.0054$ & 1.4108 \\
lower-res & MGE+AGN       & $+0.0049\pm0.0017$ & $+0.0047\pm0.0073$ & 1.3450 \\
lower-res & S\'ersic      & $+0.2278\pm0.0154$ & $+0.0064\pm0.0472$ & 3.3245 \\
lower-res & Nuker         & $+0.0631\pm0.0133$ & $-0.0306\pm0.0349$ & 1.2955 \\
lower-res & Core-S\'ersic & $+0.0480\pm0.0139$ & $-0.0378\pm0.0350$ & 1.1625 \\
\midrule
high-res  & MGE           & $+0.0038\pm0.0037$ & $+0.0025\pm0.0038$ & 1.1940 \\
high-res  & MGE+AGN       & $-0.0008\pm0.0058$ & $-0.0003\pm0.0052$ & 0.9527 \\
high-res  & S\'ersic      & $+0.0122\pm0.0121$ & $+0.0036\pm0.0115$ & 0.9946 \\
high-res  & Nuker         & $-0.0140\pm0.0161$ & $-0.0181\pm0.0183$ & 1.4350 \\
high-res  & Core-S\'ersic & $+0.0333\pm0.0219$ & $+0.0203\pm0.0194$ & 1.4415 \\
\midrule
combined  & MGE           & $+0.0059\pm0.0025$ & $+0.0060\pm0.0032$ & 1.1032 \\
combined  & MGE+AGN       & $+0.0038\pm0.0040$ & $+0.0044\pm0.0044$ & 0.9284 \\
combined  & S\'ersic      & $-0.0030\pm0.0101$ & $-0.0136\pm0.0104$ & 1.0301 \\
combined  & Nuker         & $+0.0009\pm0.0119$ & $-0.0113\pm0.0138$ & 1.4154 \\
combined  & Core-S\'ersic & $-0.1070\pm0.0284$ & $-0.0560\pm0.0289$ & 1.4273 \\
\bottomrule
\end{tabular}
\end{table}

\section{Bayesian Evidence} \label{app:evidences}
The Bayesian evidence ranking changes significantly between observations (Table \ref{tab:evidences}). For example, for the lower-resolution data, the MGE+AGN model is ranked worst and the Nuker is ranked best; meanwhile, for the higher-resolution dataset, the MGE+AGN model is ranked best and the Nuker is ranked worst. The combined dataset chooses the Core-S\'ersic as the best model, despite it not being ranked first by either of the other datasets. Furthermore, for the combined dataset, evidence-weighted model averaging would assign approximately $99.6\%$ of the weight to the Core-S\'ersic, thereby potentially artificially shrinking the systematic uncertainty on the black hole mass.
\begin{table}[H]
\centering
\caption{Log-evidence $\Delta\ln Z$ relative to the best model of each dataset (comparable only within a dataset).}
    \label{tab:evidences}
    \setlength{\tabcolsep}{3pt}
    \begin{tabular}{l ccc}
      \toprule
      Model         & lower-res         & high-res & combined \\
      \midrule
      MGE           & $-10.34$          & $-1.16$  & $-7.81$  \\
      MGE+AGN       & $-10.37$          & $\phantom{-}0.00$ & $-6.89$ \\
      S\'ersic      & $-2.30$           & $-0.37$  & $-5.97$  \\
      Nuker         & $\phantom{-}0.00$ & $-9.00$  & $-18.98$ \\
      Core-S\'ersic & $-0.59$           & $-2.96$  & $\phantom{-}0.00$ \\
      \bottomrule
    \end{tabular}
\end{table}
\section{Additional figures} \label{app:figures}
\begin{figure}[H]
    \centering
    \includegraphics[width=0.7\linewidth]{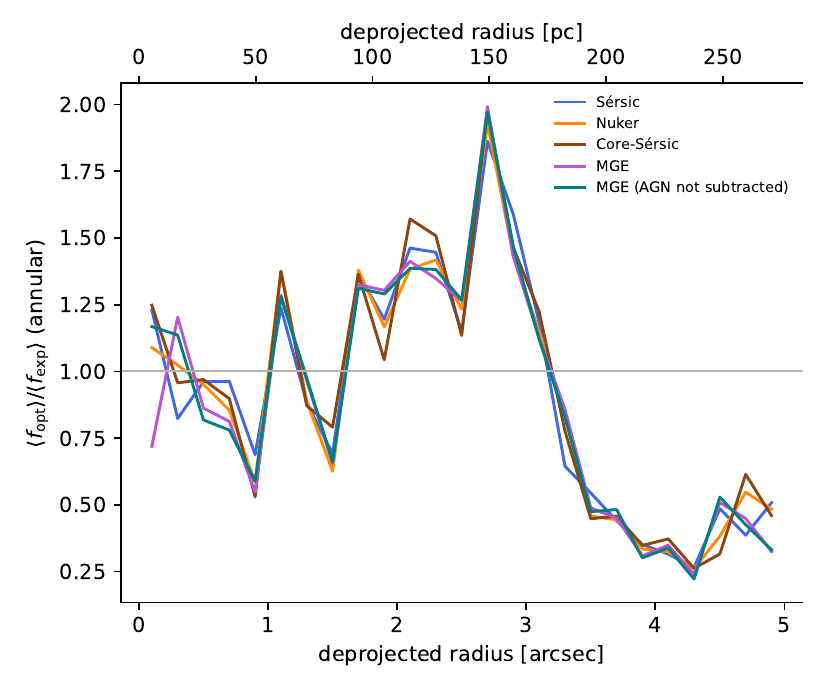}
    \caption{Discrepancy residual (lower-resolution data): azimuthally averaged ratio of the refit $\mathbf{f}_\mathrm{opt}$ to the exponential emissivity $\mathbf{f}_\mathrm{exp}$, one curve per stellar model.}
    \label{fig:flux_corrected}
\end{figure}

\begin{figure}[H]
\centering
\includegraphics[width=0.7\linewidth]{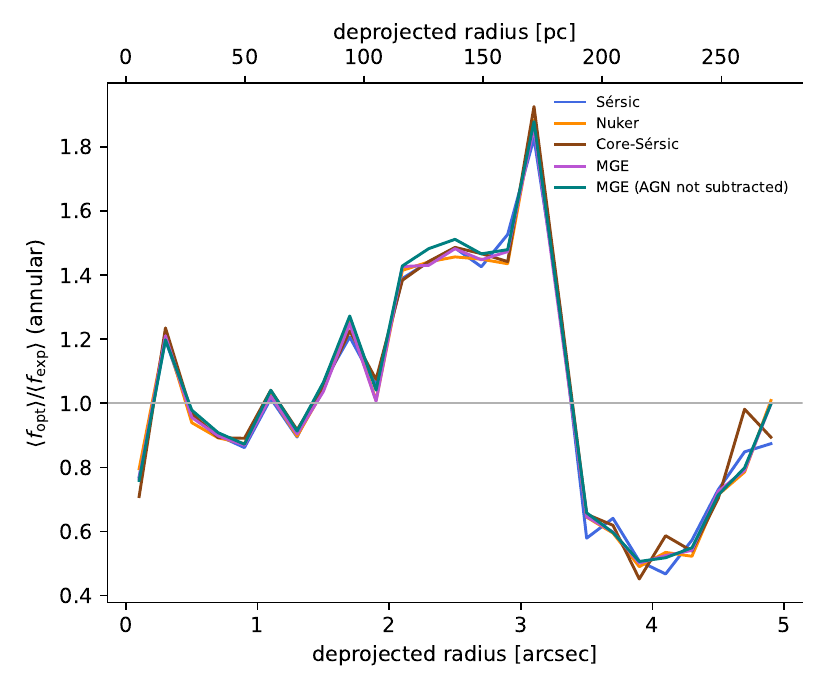}
\caption{As Figure~\ref{fig:flux_corrected}, for the combined-resolution dataset. The equivalent plot for the high-resolution dataset is nearly identical.}
\label{fig:flux_corrected_combined}
\end{figure}

In this section, we provide a visualization of the impact of re-fitting the gas emissivity according to Appendix \ref{app:refit}. In Figures \ref{fig:flux_corrected} and \ref{fig:flux_corrected_combined}, we demonstrate how the inferred gas emissivity changes as a function of radius after re-fitting: the corrected gas distribution shows stronger emission for distances near 3'' and reduced emission for distances above 3''. 

Similarly, we plot the data-model residuals before and after re-fitting the gas emissivity to demonstrate the improvement in fit. We plot the residuals using position-velocity diagrams (PVDs). A PVD is produced by extracting (from the spectral cube, whose dimensions are position, position, velocity) a slice in position situated along the kinematic major axis of the galaxy. The resulting PVD plots the position along the slice on the x axis and the velocity on the y axis. This allows simple visualization of the main kinematic features of a 3-dimensional data cube. The gas emissivity correction brings the PVD residuals down to the noise floor for all stellar mass models.

\begin{figure}[H]
\centering
\includegraphics[width=\linewidth]{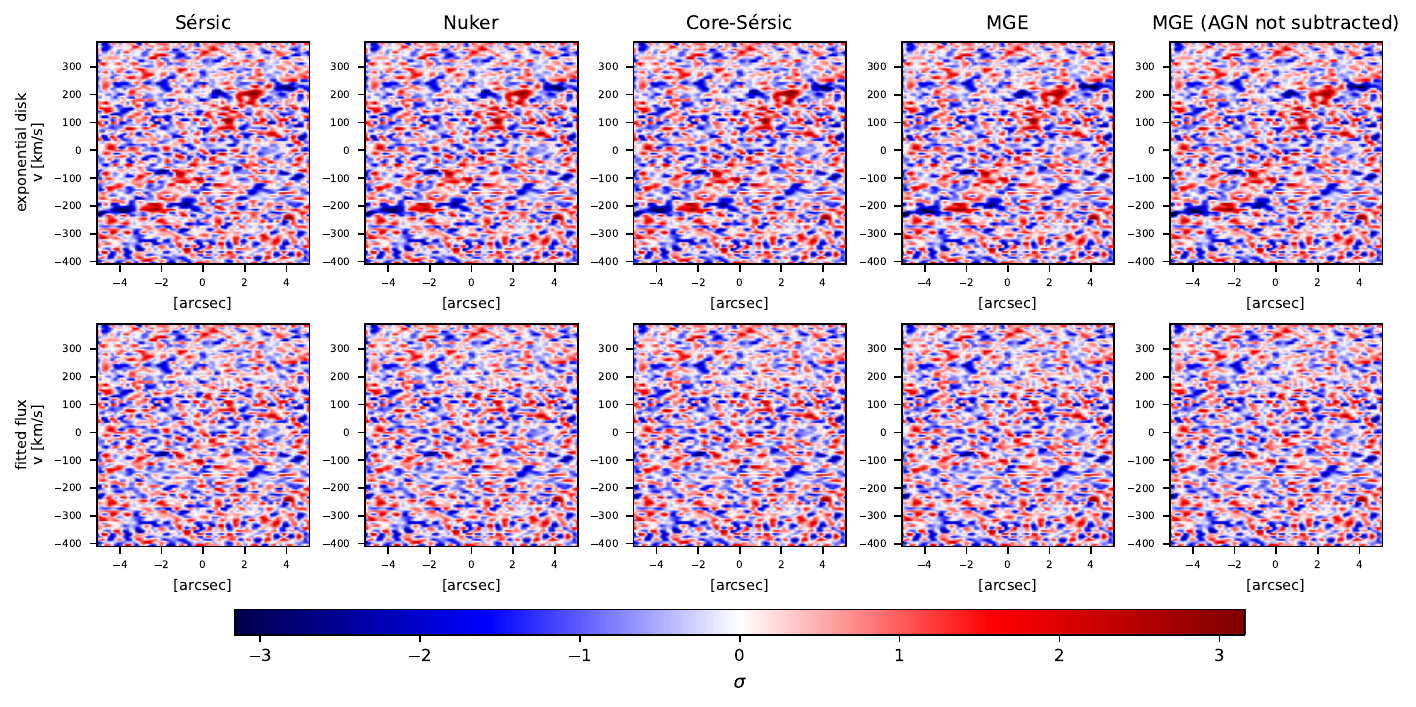}
\caption{Position--velocity residuals (data minus model, in units of the empirical noise) along the kinematic major axis of the lower-resolution dataset for the five stellar mass models, before (top row) and after (bottom row) replacing the exponential emissivity by the refit of Eq.~\eqref{eq:flux_refit}. The velocity-symmetric residual structures, identical for all stellar models, are removed by the emissivity correction, confirming that they arise from the shared emissivity model rather than from the stellar parametrizations.}
\label{fig:pvd_before_after}
\end{figure}

\section{Datasets, models and inference settings}
\label{app:settings}

\paragraph{Data.}
Both observations target the $^{12}$CO(2--1) line of NGC~4697 with the ALMA 12-m array in a compact (lower-resolution) and an extended (high-resolution) configuration; the combined dataset concatenates their visibilities. After calibration and continuum subtraction, the visibilities are regridded to a common spectral grid of 80 channels of $10\,\mathrm{km\,s^{-1}}$, Hermitian-augmented, and gridded onto a regular Fourier grid with non-overlapping cells using a Kaiser--Bessel window, so that each measured visibility contributes to exactly one cell and no inter-cell covariance is introduced. Per-cell uncertainties are estimated from the empirical weighted scatter within densely populated cells and from time-differenced per-baseline noise estimates in sparse cells. Only the non-redundant Hermitian half-plane enters the likelihood, so each independent complex visibility is counted once. Table~\ref{tab:dataset_settings} lists the resulting data volumes.

\begin{table}[H]
\centering
\caption{Dataset-dependent numerical settings of the simulator and likelihood.}
\label{tab:dataset_settings}
\small
\begin{tabular}{lccc}
\toprule
Setting & Lower resolution & High resolution & Combined \\
\midrule
Fourier grid side, $N_{uv}$ & $135$ & $1097$ & $1097$ \\
Fourier cell size, $\Delta u$ ($\lambda$) & $10313$ & $10313$ & $10313$ \\
Image pixel scale (arcsec) & $0.1481$ & $0.01823$ & $0.01823$ \\
Longest occupied gridded baseline ($\lambda$) & $4.86\times10^{5}$ & $3.85\times10^{6}$ & $3.85\times10^{6}$ \\
Model image grid, $N_\mathrm{px}$ & $69$ & $549$ & $549$ \\
Occupied half-plane cells, $N_\mathrm{cell}$ & $51\,219$ & $776\,898$ & $803\,134$ \\
Number of likelihood terms, $N=2N_\mathrm{cell}$ & $102\,438$ & $1\,553\,796$ & $1\,606\,268$ \\
\bottomrule
\end{tabular}
\end{table}

\paragraph{Simulator.}
The simulator evaluates the circular velocity of an oblate axisymmetric stellar mass distribution represented as a Multi-Gaussian Expansion \citep{Cappellari_2002a} plus a Plummer-softened point mass $\mbh$, renders a thin gas disc of exponential emissivity with a Gaussian line-of-sight velocity dispersion into a spatially and spectrally oversampled cube by a deterministic inverse mapping from sky pixels to the disc plane, and transforms the cube to model visibilities (primary-beam attenuation, flux normalization, zero-padding, the image-plane taper of the gridding window, FFT and sampling of the occupied cells). The imaging-free profiles (S\'ersic, Nuker, Core-S\'ersic) are converted to Multi-Gaussian Expansions at every likelihood evaluation by a precomputed linear projection onto a fixed bank of 64 Gaussians (reconstruction error $\lesssim0.1$ per cent), which makes the entire forward model differentiable via its implementation in \texttt{PyTorch} \citep{Paszke_2019}; the Jacobians used to cross-check Eq.~\eqref{eq:influence_mc} against the Fisher form for the imaging-based models are obtained by forward-mode automatic differentiation through this pipeline. Full details of the data reduction, simulator and priors are given in \companiont{}.

\paragraph{Parameters and priors.}
All models share nine parameters with uniform priors: the black hole mass, $\log_{10}(\mbh/\msun)\sim\mathcal{U}(2.0,9.8)$; the gas-disc scale length, inclination, position angle, gas velocity dispersion, systemic velocity, two kinematic-centre offsets and the total line flux. The imaging-based models add a single mass-to-light ratio; the imaging-free models add the shape and normalization parameters of their profile (S\'ersic: index, effective radius, normalization; Nuker: inner and outer slopes, break radius, transition sharpness, normalization; Core-S\'ersic: index, core slope, break and effective radii, transition sharpness, normalization) and a shared intrinsic axis ratio, with priors spanning the ranges observed in galaxy photometry. Tables \ref{tab:fixed_shared} through \ref{tab:prior_core_sersic} give the full parameter settings and priors for each model.

\begin{table}
	\centering
	\caption{Quantities held fixed in all dynamical models and datasets.}
	\label{tab:fixed_shared}
	\begin{tabular}{llc}
		\hline
		Quantity & Symbol & Value \\
		\hline
		\multicolumn{3}{l}{\emph{Physical and observational}}\\
		Distance                        & $D$                  & $11.4$~Mpc \\
		Scale                           & --                   & $55.27$~pc~arcsec$^{-1}$ \\
		Line rest frequency             & $\nu_0$              & $230.538$~GHz \\
		Antenna diameter                & $D_\mathrm{dish}$    & $12$~m \\
		Primary-beam FWHM coefficient   & --                   & $1.13\,\lambda/D_\mathrm{dish}$ \\
		SMBH softening length           & $r_\mathrm{soft}$    & $0.01$~pc \\[2pt]
		\multicolumn{3}{l}{\emph{Circular-velocity quadrature}}\\
		Quadrature nodes                & $n_\mathrm{quad}$    & $128$ \\
		Rotation-curve lookup radii     & $n_R$                & $4096$ \\[2pt]
		\multicolumn{3}{l}{\emph{Profile-to-MGE projection}}\\
		Gaussian basis size             & $K$                  & $64$ \\
		Radial sampling points          & $N_R$                & $100$ \\
		Radial sampling range           & --                   & $1$~pc--$10$~kpc \\
		Initial ridge parameter         & $\lambda$            & $10^{-7}$ \\[2pt]
		\multicolumn{3}{l}{\emph{Cube rendering}}\\
		Velocity sub-channels per pixel & $K$                  & $5$ \\
		Uncertainty inflation factor    & --                   & $1$ (none) \\[2pt]
		\multicolumn{3}{l}{\emph{Sampler (\texttt{Nautilus})}}\\
		Live points                     & $n_\mathrm{live}$    & $2000$ \\
		Networks per bound              & --                   & $4$ \\
		Target effective sample size    & $n_\mathrm{eff}$     & $10^4$ \\
		Random seed                     & --                   & $42$ \\
		\hline
	\end{tabular}
\end{table}

\begin{table*}
	\centering
	\caption{Priors on the free parameters shared by all dynamical models.}
	\label{tab:shared_priors}
	\begin{tabular}{llll}
		\hline
		Parameter & Symbol & Sampled quantity & Prior \\
		\hline
		Black hole mass              & $M_\mathrm{BH}$        & $\log_{10}\left(M_\mathrm{BH}/\msun\right)$      & $\mathcal{U}(2.00,\,9.80)$ \\
		Gas disk scale length        & $R_\mathrm{gas}$       & $R_\mathrm{gas}$ [pc]                            & $\mathcal{U}(0,\,552.7)$ \\
		Inclination                  & $\iota$                & $\iota$ [deg]                                    & $\mathcal{U}(72,\,89)$ \\
		Position angle               & $\phi$                 & $\phi$ [deg]                                     & $\mathcal{U}(0,\,359.9)$ \\
		Gas velocity dispersion      & $\sigma_\mathrm{gas}$  & $\sigma_\mathrm{gas}$ [km/s]                     & $\mathcal{U}(0.1,\,30)$ \\
		Systemic velocity            & $v_\mathrm{sys}$       & $v_\mathrm{sys}$ [km/s]                          & $\mathcal{U}(1220,\,1260)$ \\
		Kinematic centre offset (RA) & $x_0$                  & $x_0$ [arcsec]                                   & $\mathcal{U}(-1,\,1)$ \\
		Kinematic centre offset (Dec.)& $y_0$                 & $y_0$ [arcsec]                                   & $\mathcal{U}(-1,\,1)$ \\
		Total line flux              & $F$                    & $F$ [Jy km/s]                                    & $\mathcal{U}(1,\,50)$ \\
		\hline
	\end{tabular} \\
	{{\sl Notes.} $\mathcal{U}(a,b)$ denotes a uniform distribution on $[a,b]$. The black hole mass prior corresponds to $M_\mathrm{BH}        \in [10^{2},\,6.3\times10^{9}]~\msun$. The gas disk scale length prior corresponds to $R_\mathrm{gas} \in [0,\,10]\arcsec$ at the
         adopted distance. The offsets $x_0$ and $y_0$ are measured with respect to the phase centre of the observations (in ICRS coordinates, RA 12:48:35.8990, Dec -05:48:02.4210). The position angle is treated as a periodic parameter by the sampler.}
\end{table*}

\begin{table}
	\centering
	\caption{Parameters of the optically-derived MGE stellar mass model for the model which retains the innermost Gaussian containing the AGN signal (``MGE with AGN''). For the model with the innermost (AGN) Gaussian removed (``MGE''), we remove the first row from the table below.}
	\label{tab:mge_agn}
	\begin{tabular}{ccccc}
		\hline
		$i$ & $\log_{10}\Sigma_i$ & $\sigma_i$ & $\sigma_i$ & $q'_i$ \\
		    & ($\mathrm{L}_\odot\,\mathrm{pc}^{-2}$) & (arcsec) & (pc) & \\
		\hline
		1 & $5.579$ & $0.052$  & $2.9$   & $0.932$ \\
		2 & $4.783$ & $0.187$  & $10.3$  & $1.000$ \\
		3 & $4.445$ & $0.456$  & $25.2$  & $0.863$ \\
		4 & $4.239$ & $0.986$  & $54.5$  & $0.726$ \\
		5 & $4.061$ & $2.328$  & $128.7$ & $0.541$ \\
		6 & $3.698$ & $4.624$  & $255.6$ & $0.683$ \\
		7 & $3.314$ & $6.442$  & $356.0$ & $0.318$ \\
		8 & $3.538$ & $15.524$ & $858.0$ & $0.589$ \\
		\hline
	\end{tabular}
\end{table}

\begin{table}
	\centering
	\caption{Priors on the free parameters of the S\'ersic stellar mass model.}
	\label{tab:prior_sersic}
	\begin{tabular}{llc}
		\hline
		Parameter & Sampled quantity & Prior \\
		\hline
		S\'ersic index        & $n$                                                                          & $\mathcal{U}(0.5,\,10)$ \\
		Effective radius      & $\log_{10}\left(R_\mathrm{e}/''\right)$                                & $\mathcal{U}(-2.00,\,2.26)$ \\
		Normalisation         & $\log_{10}\left(\Sigma(R_0)/\msun\,\mathrm{pc}^{-2}\right)$                  & $\mathcal{U}(1.0,\,5.0)$ \\
		Intrinsic axis ratio  & $q$                                                                          & $\mathcal{U}(0.01,\,0.99)$ \\
		\hline
	\end{tabular} \\
	{{\sl Notes.} The effective-radius prior corresponds to
	$R_\mathrm{e} \in [0.01,\,180.9]\arcsec$, i.e.\ $[0.55~\mathrm{pc},\,10~\mathrm{kpc}]$
	at the adopted distance. $R_0=1\arcsec$. The shared parameters of Table
	\ref{tab:shared_priors} apply in addition.}
\end{table}
 
\begin{table}
	\centering
	\caption{Priors on the free parameters of the Nuker stellar mass model.}
	\label{tab:prior_nuker}
	\begin{tabular}{llc}
		\hline
		Parameter & Sampled quantity & Prior \\
		\hline
		Transition sharpness  & $\alpha$                                                                     & $\mathcal{U}(0.1,\,10)$ \\
		Inner slope           & $\gamma$                                                                     & $\mathcal{U}(0.001,\,1.2)$ \\
		Slope difference      & $\gamma-\beta$                                                               & $\mathcal{U}(-3.0,\,-0.3)$ \\
		Break radius          & $\log_{10}\left(R_\mathrm{b}/''\right)$                                & $\mathcal{U}(-2.00,\,1.48)$ \\
		Normalisation         & $\log_{10}\left(\Sigma(R_0)/\msun\,\mathrm{pc}^{-2}\right)$                  & $\mathcal{U}(1.0,\,5.5)$ \\
		Intrinsic axis ratio  & $q$                                                                          & $\mathcal{U}(0.01,\,0.99)$ \\
		\hline
	\end{tabular} \\
	{{\sl Notes.} The break-radius prior corresponds to
	$R_\mathrm{b} \in [0.01,\,30]\arcsec$, i.e.\ $[0.55,\,1658]$~pc at the
	adopted distance. The outer slope follows as
	$\beta = \gamma - (\gamma-\beta)$ and therefore satisfies
	$\gamma + 0.3 \le \beta \le \gamma + 3$ by construction. $R_0=1\arcsec$. The
	shared parameters of Table \ref{tab:shared_priors} apply in addition.}
\end{table}
 
\begin{table*}
	\centering
	\caption{Priors on the free parameters of the Core-S\'ersic stellar mass
	model.}
	\label{tab:prior_core_sersic}
	\begin{tabular}{llc}
		\hline
		Parameter & Sampled quantity & Prior \\
		\hline
		S\'ersic index        & $n$                                                          & $\mathcal{U}(2,\,11)$ \\
		Transition sharpness  & $\alpha$                                                     & $\mathcal{U}(1.5,\,10)$ \\
		Core slope            & $\gamma$                                                     & $\mathcal{U}(0.001,\,1.2)$ \\
		Break radius          & $\log_{10}\left(R_\mathrm{b}/''\right)$                & $\mathcal{U}(-2.00,\,1.48)$ \\
		Effective radius      & $\log_{10}\left(R_\mathrm{e}/''\right)$                & $\mathcal{U}\left(\log_{10}\left[1.1\,R_\mathrm{b}/''\right],\,2.26\right)$ \\
		Normalisation         & $\log_{10}\left(\Sigma(R_0)/\msun\,\mathrm{pc}^{-2}\right)$  & $\mathcal{U}(1.0,\,5.5)$ \\
		Intrinsic axis ratio  & $q$                                                          & $\mathcal{U}(0.01,\,0.99)$ \\
		\hline
	\end{tabular} \\
	{{\sl Notes.} The break-radius prior corresponds to
	$R_\mathrm{b} \in [0.01,\,30]\arcsec$, i.e.\ $[0.55,\,1658]$~pc at the
	adopted distance. The effective radius is sampled under a prior conditional
	on $R_\mathrm{b}$, uniform in $\log_{10}R_\mathrm{e}$ between
	$1.1\,R_\mathrm{b}$ and $180.9\arcsec$ ($10$~kpc at the adopted distance).
	$R_0=1\arcsec$. The shared parameters of Table \ref{tab:shared_priors} apply
	in addition.}
\end{table*}

\paragraph{Sampling and compute resources.}
Each posterior is sampled with \texttt{Nautilus} \citep{Lange_2023} using 2000 live points, four neural networks per bound, a target of $10^4$ effective posterior samples and a fixed random seed, discarding the exploration phase; live points are distributed over the GPUs of a single node, each holding an independent copy of the simulator. The forward model runs in single precision with the $\chi^2$ accumulated in double precision. Influence maps require $n_s=512$ additional forward evaluations per model and dataset; the null calibration requires $n_\mathrm{mock}=200$ constrained refits per model and dataset.

All nested-sampling fits were run on national academic HPC clusters (nodes with 4 $\times$ A100 40 GB; nodes with 8 $\times$ H200 141 GB). Data products total roughly 0.7 TB on local storage (roughly 640 GB raw and gridded visibilities; roughly 73 GB posterior and analysis products). Each reported fit took 0.6–12.9 node-hours of wall-clock time depending on model complexity ($2.1\times10^5$–$3.7\times10^6$ likelihood evaluations); counting additional models that were tested, the fits total about 63 node-hours (about 108 A100-h + roughly 287 H200-h). Null-calibration mocks (which repeat the flux solve 200 times per dataset) add about 20 A100-hours, and Jacobians, flux re-optimization, and influence-map harvesting add $\approx 10$ GPU-hours, for $\sim 425$ GPU-hours of cluster compute in total.

The complete research project used substantially more compute than the reported experiments---about 5,000 additional GPU-hours---on 
preliminary and failed campaigns, including the following: MALA/parallel-tempering MCMC experiments ($\approx 1,800$ GPU-h), an UltraNest production campaign that failed due to a float32 likelihood-quantization bug and its float64 rerun ($\approx 1,200$ GPU-h), sampler benchmarking and prior/parametrization studies ($\approx $1,100 GPU-h), and a superseded nested-sampling generation replaced after a code review ($\approx700$ GPU-h).

LLM-based tools were used for project organization/planning, language editing, code debugging assistance, and drafting code (supervised by the authors). All resulting code, derivations, citations, figures, and prose were independently reviewed and validated by the authors. LLMs were not components of the inference method evaluated in this work. All generative AI code and text have been verified by the authors and have been saved with version control to ensure reproducibility. The authors of this work accept final responsibility for all published results.

\end{document}